\documentclass[
]{ceurart}

\usepackage{listings}
\usepackage{caption}
\usepackage{algorithm}
\usepackage{algpseudocode}
\begin{document}

\copyrightyear{2022}
\copyrightclause{Copyright for this paper by its authors.
  Use permitted under Creative Commons License Attribution 4.0
  International (CC BY 4.0).}

\conference{AIQxQIA 2026: International Workshop on AI for Quantum and Quantum for AI | co-located with AIxIA 2026, Perugia, Italy}

\title{Photometric Redshift Estimation with Hybrid Quantum Neural Networks}

\tnotemark[1]

\author[1]{Giuseppe Sarracino}[%
orcid=0000-0001-9512-1961,
email=giuseppe.sarracino@inaf.it
]
\cormark[1]
\address[1]{INAF-Osservatorio Astronomico di Roma, Via Frascati 33, Monteporzio Catone, Roma, 00078, Italy}

\author[1]{Vincenzo Fabrizio Cardone}[%
orcid=0000-0002-2079-7438,
]

\author[2]{Irene Graziotti}[%
]
\address[2]{INAF-Osservatorio Astronomico di Capodimonte, Via Moiariello 16, 80131, Napoli, Italy}

\author[2]{Crescenzo Tortora}[%
orcid=0000-0001-7958-6531,
]

\author[1]{Roberto Scaramella}[%
orcid=0000-0003-2229-193X,
]

\author[3]{Andrea Bulgarelli}[%
orcid=0000-0001-6347-0649,
]

\address[3]{INAF-Osservatorio di astrofisica e scienza dello spazio, Via Pietro Gobetti 93/3, Bologna, 40129, Italy}

\author[4]{Carlo Burigana}[%
orcid=0000-0002-3005-5796,
]
\address[4]{INAF-Istituto di Radioastronomia, Via Piero Gobetti 101, Bologna, 40129, Italy}

\author[5]{Farida Farsian}[%
orcid=0009-0004-5252-1737,
]
\address[5]{INAF-Osservatorio Astronomico di Catania, Via S. Sofia 78, Catania, 95123, Italy}

\author[3]{Massimo Meneghetti}[%
orcid=0000-0003-1225-7084,
]

\author[6]{Giuseppe Murante}[%
orcid=0000-0002-5155-130X,
]
\address[6]{INAF-Osservatorio Astronomico di Trieste, Via Tiepolo 11, Trieste, I-34131, Italy}

\author[3]{Niccolò Parmiggiani}[%
orcid=0000-0002-4535-5329,
]

\author[5]{Alessandro Rizzo}[%
orcid=0009-0003-4341-2988,
]

\author[5]{Francesco Schillirò}[%
orcid=0000-0001-5106-2277,
]

\author[1]{Vincenzo Testa}[%
orcid=0000-0003-1033-1340,
]

\author[4]{Tiziana Trombetti}[%
orcid=0000-0001-5166-2467,
]

\author[5]{Fabio Roberto Vitello}[%
orcid=0000-0003-2203-3797,
]

\cortext[1]{Corresponding author.}

\begin{abstract}
  A Hybrid Quantum Neural Network (HQNN) is applied to recover the photometric redshift of a galaxy dataset built using data from two astronomical surveys, the Kilo-Degree Survey (KIDS) and the VISTA Kilo-Degree Infrared Galaxy (VIKING), which provide data with coverage from the optical to the near-infrared electromagnetic spectrum. The results are then compared with those obtained by a classical neural network. We see that HQNN consistently recovers the reference photometric redshift with a high statistical fidelity. Furthermore, both the $R^2$ and the MSE found by the HQNN are consistently better than those found by the classical counterpart (albeit consistent in the error range), proving the worth of further investigating these novel technologies for this particular astrophysical application.
\end{abstract}


\maketitle

\section{Introduction}

Quantum Computing (QC) is a novel technology merging concepts of Computer Science and Quantum Mechanics \cite{Yanofsky_2007}. In recent years, an impressive technological advancement has brought us to the first proper quantum hardware \cite{Johnson_2011, Kim_2023, Massarotti_2023}, so that we are currently in what is called the Noisy Intermediate Scale Quantum (NISQ) era \cite{Preskill_2018}. The scientific community is now looking for applications where we reach the so-called ``quantum utility", which is a general improvement of the QC results with respect to the classical counterpart. In particular, we focus on possible applications of QC to cosmology and astrophysics \cite{Sarracino_2025, Sarracino_2026, farsian_2025, farsian_2025_b, Farsian_2026, Rizzo_2026, Bulgarelli_2026}. Indeed, we live in an era where new astrophysical missions and instruments \cite{Abell_2009, Maartens_2015, Scaramella_2022} will bring us impressive datasets, which require strategies like machine learning \cite{Angora_2020} or a high-performance computing (HPC) framework to be properly analyzed \cite{Taffoni_2020}. In this context, we also test QC as a novel, alternative strategy.

In this work, we present the first tests on a Hybrid Quantum Neural Network (HQNN, i.e. the quantum component is the circuit itself, while the gradient descent is classical, as we will detail) to infer the photometric redshift (i. e., an estimate of the recession velocity of an astronomical object performed without using its spectrum, differently from the spectroscopic redshift) of a sample of galaxies selected from the Kilo-Degree Survey (KIDS) \cite{Kuijken_2019} and the VISTA Kilo-Degree Infrared Galaxy (VIKING) Survey \cite{Edge_2013}. This dataset is composed of approximately 125500 galaxies with KIDS+VIKING photometry chosen with redshift $z<0.5$ \cite{Humphrey_2025}, with the main aim of testing astrophysical models and algorithms, such as machine learning methods applied to astrophysics \citep{Cavuoti_2015, Cavuoti_2017} principally involving galaxy evolution. The dataset has been built by extracting, among other parameters, the photometric redshift and nine-band Gaussian Aperture and PSF (GaAP) magnitudes \cite{Kuijken_2019} ($u,g,r,i, Z, Y, J, H, K_s$). More specifically, the photometric redshifts have been derived by using the so-called Bayesian Photometric Redshift (BPZ, \cite{Benitez_2000}) technique from the observations; for more details, see \cite{Kuijken_2019}. We use these photometric redshift for our analysis, as a reference point. In particular, the core idea of this work is to use the color bands of the datasets (derived from the observed magnitudes) to infer the photometric redshift from the HQNN after a pre-processing step, compare the results with the reference photometric redshifts provided by the dataset, and see how it works in comparison with a classical Neural Network (NN). In Section 2, we present our HQNN, in Section 3 we present the results, while conclusions are shown in Section 4.

\section{The Algorithm}

\begin{algorithm} \label{Alg_HQNN}
\caption{Hybrid Quantum Neural Network (HQNN) for photometric redshift estimation}
\begin{algorithmic}[1]

\State Split data into training, validation, and test sets
\State Standardize inputs and transform the target redshift
\State Apply \textit{PCA} and rescale features to $[-\pi,\pi]$
\State Build hybrid model: classical linear layer + variational quantum circuit
\For{each epoch}
    \State Shuffle training data
    \For{each mini-batch}
        \State Compute predictions
        \State Evaluate \textit{weighted MAE} loss
        \State Update trainable parameters
    \EndFor
    \State Compute weighted MAE loss on the validation set
    \If{validation loss improves}
        \State Save the best model
    \Else
        \State Increase \textit{patience} counter
    \EndIf
    \If{\textit{patience} exceeded}
        \State Stop training
    \EndIf
\EndFor
\State Reload best model
\State Predict on test set and invert target transformations
\end{algorithmic}
\end{algorithm}

Our HQNN algorithm is summarized in Alg. 1. Here, we start from seven photometric colours, three derived from the KiDS bands, while four from the VIKING bands. We first split the data in $80\%$ training set, $10\%$ validation set, and $10\%$ test set. Then, we standardize the training  photometric redshift set by using a quantile transformation with a normal output distribution, which is then rescaled to $[-1,1]$ interval range using a MinMax transformation, which has been done because it is the interval range in which the output of the quantum observable is defined. We then perform a Principal Component Analysis (PCA) to compress the seven starting features into 5 principal components, maintaining around $96.7 \%$ of the total variance. Then, the rescaling to $[-\pi,\pi]$ is performed so that the features can be properly translated into the parametrized quantum gates. In the classical preprocessing part of HQNN, after a linear transformation, defined so that each Qubit may depend on a combination of the principal components rather than just one, we define as an activation function an hyperbolic tangent, mapping the output to (-1,1). Then, these are multiplied to the previous $[-\pi,\pi]$ range, to be used as the angles in the rotation gates of the quantum part of HQNN. For each computation, the training set is divided into batches of 200 data points, with the model updated after each of those to decrease the computational cost of a single step. We stress that the preprocessing transformations are fitted on the training set and then applied to the test and validation ones.

\begin{figure} 
  \centering
  \includegraphics[width=\linewidth-305pt]{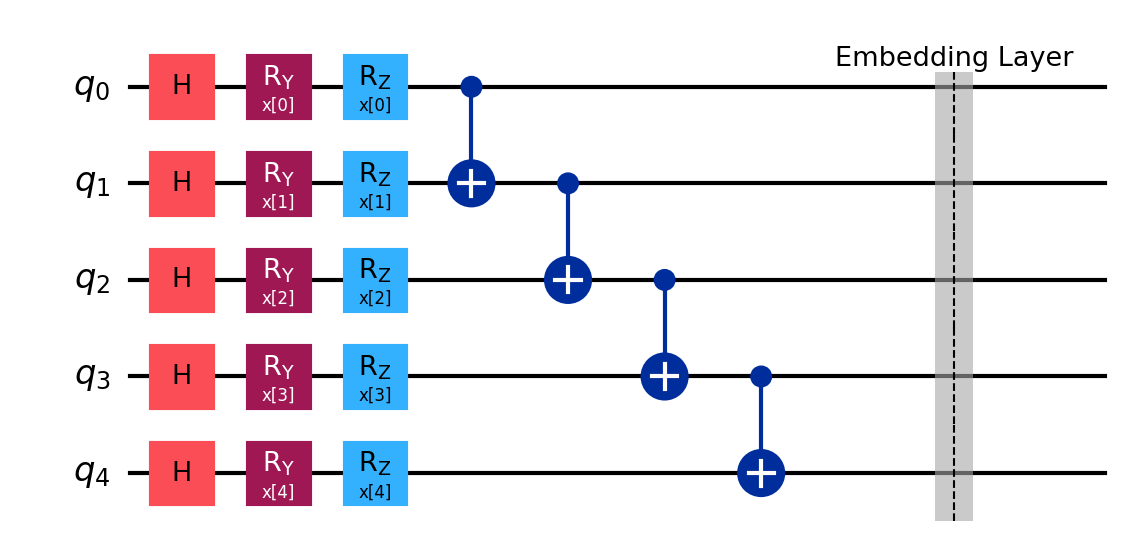}
  \includegraphics[width=\linewidth-200pt]{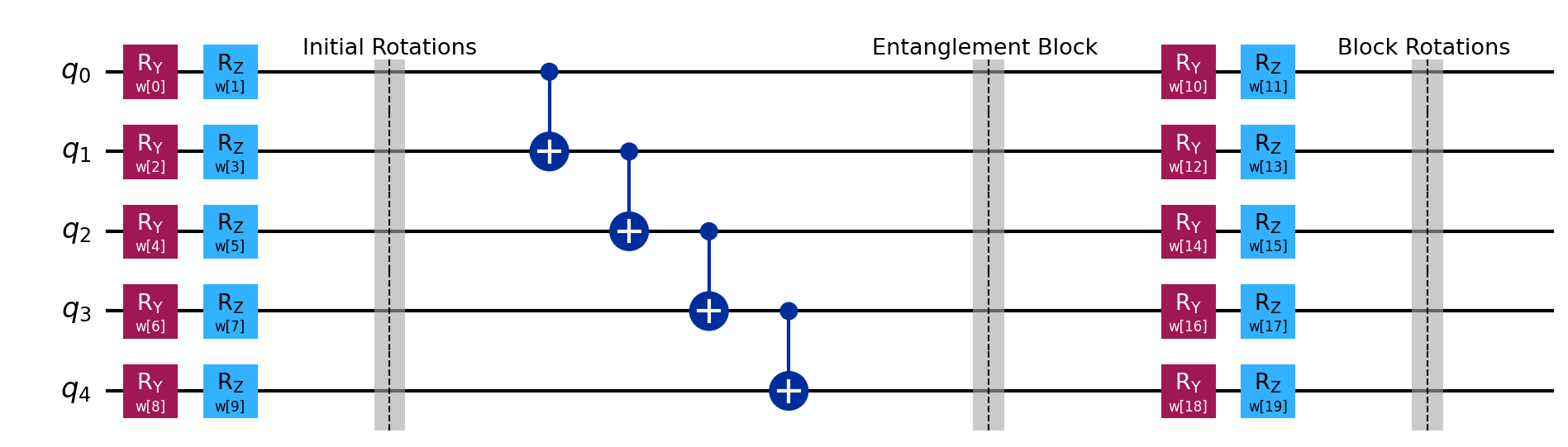}
  \caption{The Ansatz used for our HQNN. On the left panel, the initial embedding layer to encode the classical data inside the quantum circuit; on the right panel, the initial trainable rotation block followed by one of the five repetitions composing the total network.}
  \label{Fig_Quantum_Circuits}
\end{figure}

We have chosen a particular ansatz for both the form of the circuit and the observable to be measured.  More specifically, our quantum circuit is made up of 5 qubits (one for each of the five PCA components), and is composed of one embedding layer,  where we have a Hadamard gate applied to all qubits followed by two sets of rotation gates whose rotations depend on the input parameter and a series of CNOT connecting them, and is needed to encode the classical scaled data inside the quantum circuit, and 5 repetition blocks made up of two parametrized rotational blocks plus other entanglement sets of CNOT Gates. In total, there are 60 trainable quantum parameters in this configuration, considering also an initial rotational block before the repetition ones. This is displayed in Fig. \ref{Fig_Quantum_Circuits}.

The expectation value, instead, has been measured by considering the observable  $\hat O = \frac{1}{n}\sum_{j=1}^{n} Z_j,$ where $Z_j$ is the Pauli-Z acting on the qubit $j$ and the identity on all the others, while $n$ is the number of qubits. Up to now, for our quantum computations, we have used Pennylane\footnote{https://pennylane.ai/} for emulating the algorithm on classical hardware. For the implementation details, we have used the Adam optimizer built into Pytorch, with a learning rate of $10^{-2}$. The simulations have been performed using the \textit{lightning.qubit} statevector backend with the adjoint differentiation method, which computes the exact analytical gradients by leveraging on the structure of the quantum circuit, and thus no finite-shot sampling has been used. 
As for the classical part of the HQNN, the function to minimize is a weighted Mean Absolute Error (MAE), defined as:
\begin{equation} \label{MAE}
    MAE:=\frac{1}{\sum_{i=1}^Nw_i}\sum_{i=1}^Nw_i|y_i-\hat{y}_i|,
\end{equation}
where $w_i$ are weights that take into account the non-uniform redshift distribution of the dataset by being defined as inversely proportional to the number of training objects present in a given bin of the dataset, thus giving more weight to  regions with a smaller number of objects, and $y_i$ the data points. Here, N is the number of data points inside a given batch.
The \textit{patience} for our HQNN has been chosen so that we declare convergence if no improvement has been found after 10 consecutive epochs.
\section{Results}

We choose subsets of our dataset randomly and run our HQNN, performing various tests considering sets from 2000 up to 30000 data points, of which we recall that $80\%$ make the training sample (for now, we did not use the entire set for computational cost reasons). For each of these,  we have performed 10 runs to have statistics on the final metrics (for each of these, the batching and reshuffling of the dataset is done independently). As an example, in Fig. 2 we show the results for a run with 10000 data points. On the left panel, we show the comparison between the redshifts found by HQNN and the reference ones; on the right panel, instead, we show the plot of the residuals. We note how our results recover almost perfectly $z_{pred}=z_{test}$ as the best-fit relation, confirming the reliability of the analysis. HQNN also achieves an impressive  $R^2=0.970$. On the left panel we show the value for the $\sigma_{NMAD}$ (Normalized Median Absolute Deviation), defined as 
\begin{equation} \label{sigma_NMAD}
    \sigma_{NMAD}:=1.48\operatorname{median}\biggl(\frac{|y_i-\hat{y}_i|}{1+y_i}\biggr).
\end{equation}
This is a robust error metric used in regression, especially valuable when dealing with outliers, as it employs the median instead of the mean. It is also used to define the $\sigma_i$ errors shown in the left panel of the same figure.

\begin{figure} \label{Fig_One_Run_HQNN}
  \centering
  \includegraphics[width=\linewidth-80pt]{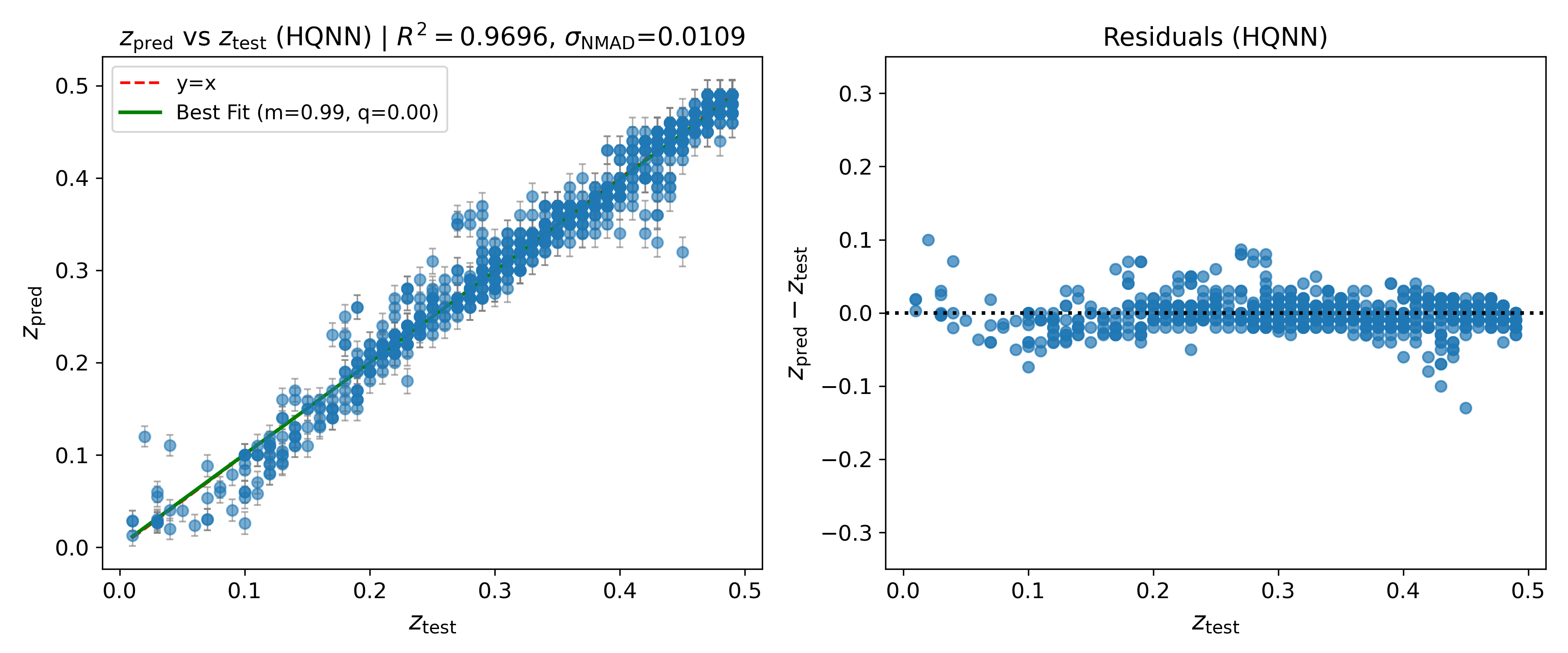}
  \caption{One exemplary result from our tests for photometric redshift derived by our HQNN in comparison with the reference redshift data from the KIDS+VIKING dataset. Here, the errors on the single points are defined as $\sigma_i= \sigma_{NMAD}(1+z_i)$, to reflect the typical error magnitude at different redshifts. On the left panel, the red line almost coincides with the green one.}
\end{figure}

\begin{figure} \label{Fig_statistics}
  \centering
  \includegraphics[width=\linewidth-260pt]{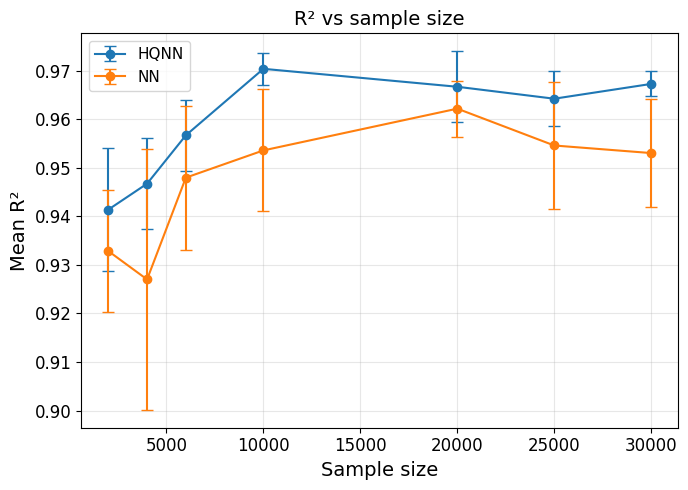}
  \includegraphics[width=\linewidth-260pt]{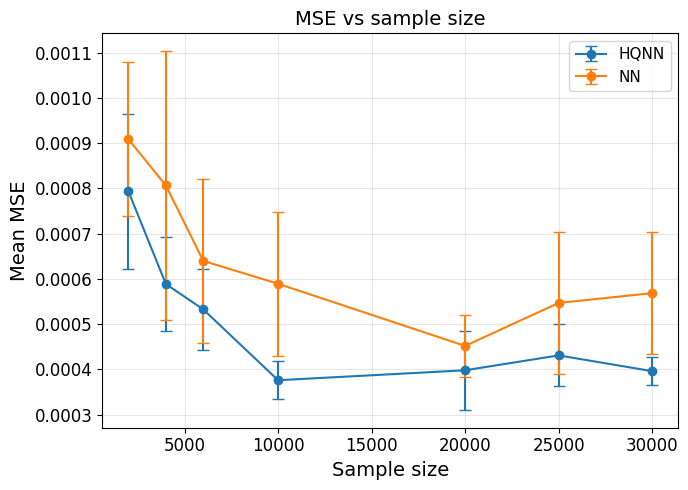}
  \includegraphics[width=\linewidth-260pt]{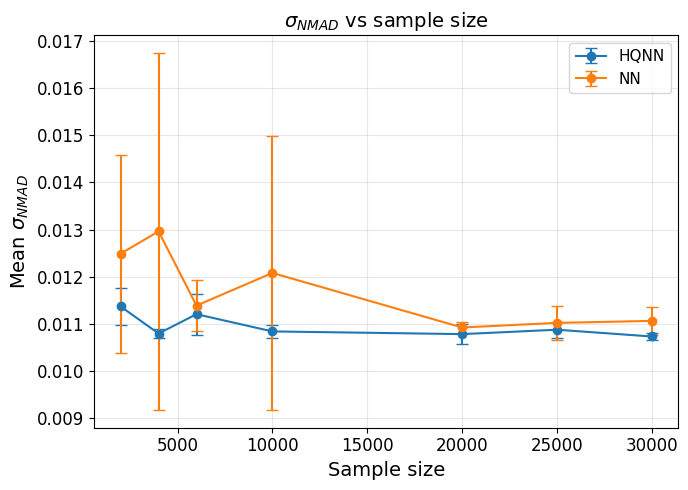}
  \caption{Summary statistics of our results comparing the performance of our HQNN to a classical NN. Here are the mean and standard deviations of the statistical indicators considered in our analysis.}
\end{figure}

In Fig. 3 we show a summary of our findings, using as a comparison a classical NN with a comparable number of parameters (59) to the quatum part of HQNN, built using two hidden layers and an input one matching the number of features used after the PCA (5), thus with an overall architecture (5-5-4-1) and using the same operations (splitting, rescaling, transformation, Adam optimizer, mini-batches strategy, WMAE loss function, and \textit{patience} stopping criterium) used for the HQNN, while using the ReLu activation function between the layers. 
Here, MSE is the Mean Squared Error, defined as $MSE=\frac{1}{N}\sum_{i=1}^N(y_i-\hat{y}_i)^2$. We note how, overall, our HQNN achieves slightly better results. More specifically, the $R^2$ value is systematically larger for HQNN, while conversely, the MSE is smaller. While the results between HQNN and NN are always consistent within the errors found by us by running the experiments 10 times each, this is still a positive outcome for HQNN. It is also worth noting how the errors on the statistical results found by HQNN are almost always smaller than the corresponding ones found by NN. While this could mainly be due to the small number of experiments performed, this is still interesting to see because it means that HQNN presents lower variability with respect to the classical counterpart, which is a point worth investigating in future experiments.

\section{Conclusions and outlooks}

We have presented the first implementation and results of an HQNN architecture to find the photometric redshift of an astrophysical dataset. The results are encouraging, because the application to the KIDS+VIKING dataset has given us convincing results both with the fidelity of the redshifts found by our algorithm and in the overall statistical evaluations we have performed, also in conjunction with a classical NN built to be as similar to our HQNN as possible. For the runtime, the classical NN is considerably faster than our HQNN. This could be because we are currently using emulators for our quantum computations \cite{Young_2023}. As a development of this research line, an outlook will be to perturb the dataset with noise on the redshift to understand if the HQNN still provides slightly better and more stable results than the classical counterpart. Moreover, we could also include other astrophysical parameters, like the stellar mass or the star formation rate. 
We also plan to test this algorithm on quantum hardware, given that the number of qubits is fairly limited for our current applications, or involving more data-rich datasets, like COSMOS2020 \cite{Weaver_2022}, which contains 1 million objects measured in more than 30 photometric bands from the UV to the infrared, as well as 487666 spectroscopic redshift measurements \cite{khostovan_2025}.

\bibliography{Main_file.bib}

\end{document}